\documentclass[sigconf]{acmart}

\usepackage{verbatim}
\usepackage{enumitem}
\usepackage{multirow}
\usepackage{color, colortbl}
\definecolor{light-gray}{gray}{0.95} 
\definecolor{LightGray}{gray}{0.95}
\usepackage{listings} 
\usepackage{xcolor} 

\AtBeginDocument{%
  }

\copyrightyear{2025}
\acmYear{2025}
\setcopyright{cc}
\setcctype{by-nc}
\acmConference[CHI '25]{CHI Conference on Human Factors in Computing Systems}{April     26-May 1, 2025}{Yokohama, Japan}
\acmBooktitle{CHI Conference on Human Factors in Computing Systems (CHI '25), April     26-May 1, 2025, Yokohama, Japan}\acmDOI{10.1145/3706598.3713815}
\acmISBN{979-8-4007-1394-1/25/04}

\newcommand{\paragraphB}[1]{\paragraph{{\textbf{#1}}}}

\begin{document}

\title[\toolkit]{\toolkit: A Toolkit for Rapid Prototyping of Interactions for Arm-based Exoskeletons }

\newcommand{\toolkit}{ExoKit}

\author{Marie Muehlhaus}
\email{muehlhaus@cs.uni-saarland.de}
\affiliation{%
  \institution{Saarland University, Saarland Informatics Campus}
  \city{Saarbrücken}
  \country{Germany}
}

\author{Alexander Liggesmeyer}
\email{liggesmeyer@cs.uni-saarland.de}
\affiliation{%
  \institution{Saarland University, Saarland Informatics Campus}
  \city{Saarbrücken}
  \country{Germany}
}


\author{Jürgen Steimle}
\email{steimle@cs.uni-saarland.de}
\affiliation{%
  \institution{Saarland University, Saarland Informatics Campus}
  \city{Saarbrücken}
  \country{Germany}
}


\renewcommand{\shortauthors}{Muehlhaus et al.}


\begin{abstract}
Exoskeletons open up a unique interaction space that seamlessly integrates users' body movements with robotic actuation. Despite its potential, human-exoskeleton interaction remains an underexplored area in HCI, largely due to the lack of accessible prototyping tools that enable designers to easily develop exoskeleton designs and customized interactive behaviors. 
We present \toolkit, a do-it-yourself toolkit for rapid prototyping of low-fidelity, functional exoskeletons targeted at novice roboticists. 
\toolkit~includes modular hardware components for sensing and actuating shoulder and elbow joints, which are easy to fabricate and (re)configure for customized functionality and wearability. To simplify the programming of interactive behaviors, we propose functional abstractions that encapsulate high-level human-exoskeleton interactions. 
These can be readily accessed either through \toolkit's command-line or graphical user interface, a Processing library, or microcontroller firmware, each targeted at different experience levels.
Findings from implemented application cases and two usage studies demonstrate the versatility and accessibility of \toolkit~for early-stage interaction design. 

\end{abstract}


\begin{teaserfigure}
  \includegraphics[width=\textwidth]{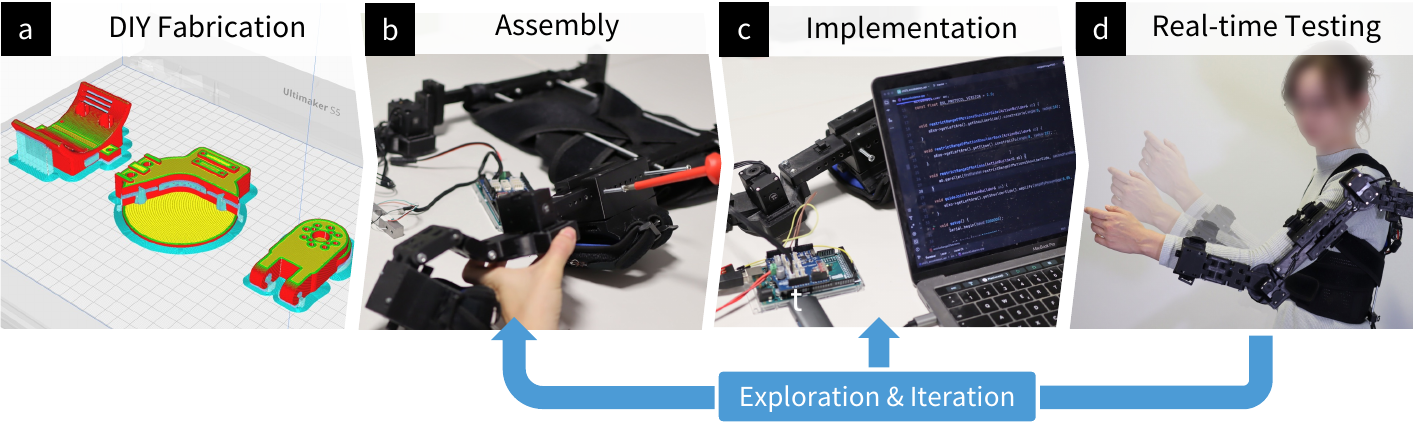}
  \caption{
    \toolkit~enables rapid prototyping of low-fidelity exoskeletons using 3D-printed and off-the-shelf components~(a). Its modular design allows easy customization of functionality and size~(b). To ease customization of the exoskeleton's interactive behavior, functional abstractions of basic functionalities and higher-level augmentation strategies are readily accessible through programming interfaces~(c). 
    The designed interaction can be tested in real-time~(d).}
  \label{fig:teaser}
  \Description{Overview of ExoKit’s workflow. The diagram consists of four sub-images, representing the stages of ExoKit’s workflow. (a) DIY Fabrication—a 3D printer software displays digital 3D models that shall be printed. (b) Assembly—a user assembles the modular, 3d printed components. (c) Implementation—a laptop connected to an Arduino is wired to the assembled hardware. A user implements some code. (d) Real-time Testing—the user wears the exoskeleton while the arm is being moved. The user can iterate and explores further alternatives, either restarting at the assembly or the implementation stage.}
\end{teaserfigure}


\maketitle

\section{Introduction}
Exoskeletons create unique opportunities for novel interactions which assist, modify or constrain physical body movement. For instance, recent work in the HCI community has leveraged hand-based or elbow-based exoskeletons for diverse applications including motion capture~\cite{gu_2016}, haptic feedback in VR~\cite{gu_2016, teng_2022}, or as a means for skill transfer between people~\cite{nishida_2022}.

Despite the great potential of exoskeletons, and despite decades of research in other communities, human-exoskeleton interaction is still a largely underinvestigated field in HCI. 
A key challenge contributing to this gap is the lack of affordable exoskeleton technologies and the lack of easily customizable and easily programmable commercial exoskeletons. The degree of expertise currently required for engineering an exoskeleton and for controlling its motion presents high barriers of entry for novices. This is further aggravated by a certain disciplinary gap in goals between HCI and robotics: while exoskeletons from robotics tend to address specific applications with sophisticated high-end technology~\cite{gull_2020,bogue_2018,kapsalyamov_2020}
, interaction designers need support for creatively exploring design opportunities in the early design phases. This requires low-fidelity, easily customizable and easy-to-iterate-on prototypes~\cite{preece_2015}.

We address this gap with \toolkit, an \textbf{open-source toolkit for rapid prototyping of low-fidelity yet functional exoskeleton prototypes}. \toolkit~is targeted at researchers, hobbyists, and makers with basic electronics and programming skills (e.g., gained through Arduino-based projects). Expertise with robotics, including their design and control, is not required\footnote{We refer to this target group as \textit{novice roboticists} for the remainder of the paper.}. \toolkit~aims to support them in creatively designing and exploring novel human-exoskeleton interactions in the early design phases. 
\toolkit~is informed by a set of considerations that we derived from the literature on HCI toolkit research, wearables and exoskeletons. 
The key benefits of \toolkit~are: (a)~\textit{ease of development for novice roboticists}, achieved with a DIY approach that uses 3D printed components, off-the-shelf servo motors, and a dedicated Arduino firmware library, and (b)~\textit{customizability in design and behavior for rapid prototyping}, achieved through modularization of the hardware and software required for prototyping human-exoskeleton interactions. The resulting building blocks can be easily exchanged or combined into new functionalities, promoting the exploratory and iterative character of interaction design for a wide range of applications. 

\toolkit's \textbf{hardware components} realize a low-fidelity exoskeleton prototype that can actuate one or two arms, with up to two active degrees-of-freedom (DoF) at the shoulder and one active DoF at the elbow. 
Arms hold promise for novel interactions in versatile application areas, such as physical motion guidance or strength augmentation; however, they present challenges in the mechanical design~\cite{tiseni_2019}. 
To tailor the prototype to the needs of diverse applications, the toolkit offers modular hardware components, which allow for hands-on reconfigurability of the exoskeleton's functionality, enable adjusting the size to accommodate various body sizes, and deploy important safety mechanisms.

\toolkit's \textbf{software library} provides functional abstractions which free the novice roboticist from detailed low-level motion control. We first identified frequently used augmentation strategies from state-of-the-art exoskeleton literature~\cite{proietti_2016,gasperina_2021} and organized them in a two-dimensional conceptual space. 
For each identified augmentation strategy, \toolkit~provides pre-implemented functions that the developer can use off-the-shelf for rapid experimentation and ideation. Amongst others, these comprise functions that interactively amplify or resist user motion, modulate the style of a motion, transfer motion from one exoskeleton to another, or guide the user's motion through real-time haptic feedback. For the ease of development, \toolkit~ provides simplified access to these functions through a command-line interface, a GUI, a Processing library, or directly through the Arduino firmware.
The firmware library enables the novice roboticist to compose the offered functional abstractions into more complex and meaningful interactions.
We make the hardware design and software libraries openly available\footnote{\url{https://github.com/HCI-Lab-Saarland/ExoKit}}.

To confirm the versatility of the toolkit for rapid prototyping, we demonstrate application cases that have been successfully realized with the framework, discuss their iterative design process and lessons learned. 
Second, we present results from two usage studies. In our first study, we collect feedback of \toolkit~in use, and discuss user's opinions on the provided functional abstractions, programmability, and wearability. With our second user study, we provide insights into how users approached \toolkit~to create their own applications, their workflows and encountered challenges.
We conclude by discussing implications for the design of human-exoskeleton interactions, limitations, and future directions.

In summary, the main contribution of this paper is \toolkit, a toolkit for rapid prototyping of human-exoskeleton interaction, targeted at novice roboticists in the early phases of interaction design. We
\begin{itemize}
    \item 
    conceptually identify functional abstractions that encapsulate relevant augmentation strategies, implemented in an Arduino firmware library. A designer can readily access these through a command-line interface, GUI, Processing library, or further customize the implemented strategies directly through the firmware.
    \item modularize the exoskeleton into hardware components that can be easily exchanged or combined into new functionalities, adjusted in size, and used to implement physical safety mechanisms. The resulting prototype can support the user with up to 3 DoF per arm.
    \item  demonstrate \toolkit's versatility and utility through application examples and two usage studies. 
\end{itemize}

We hope that this work will inspire and empower HCI researchers, designers, and makers alike to start engaging with the exciting area of human-exoskeleton interaction and exploring its potential for innovative applications.

\begin{figure*}[tb]
    \centering
    \includegraphics[width=0.78\linewidth]{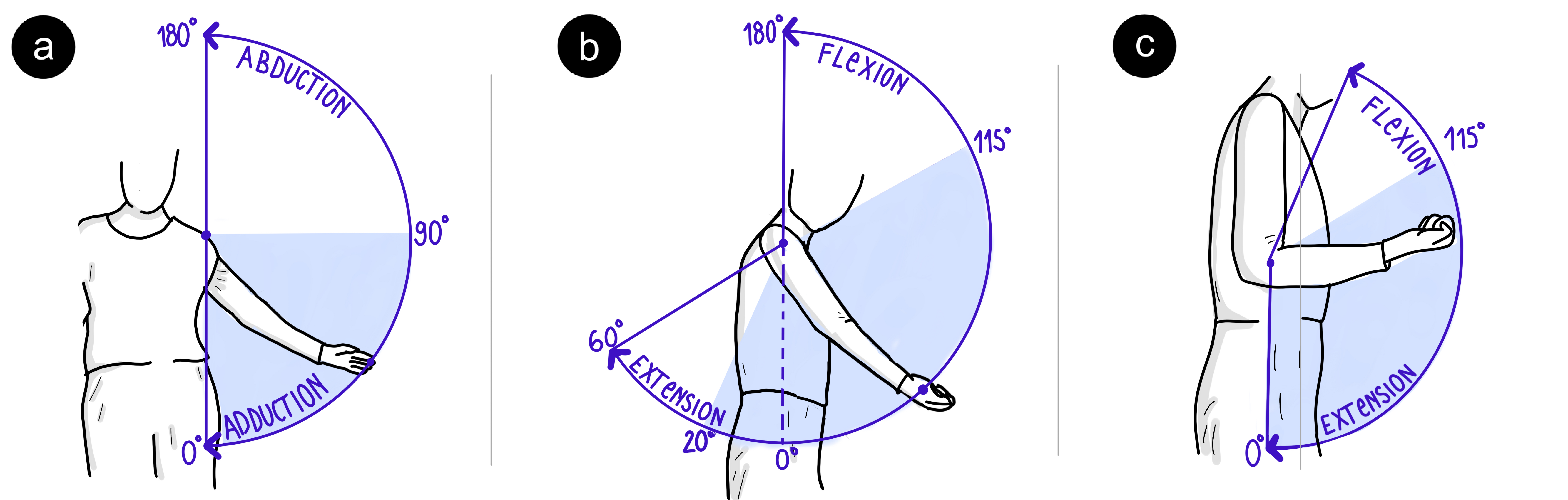}
    \caption{Degrees of freedom supported by ExoKit and the joints' ranges of motion (RoM) according to~\cite{norkin_2016}. The colored areas indicate the RoMs supported by ExoKit.}
    \label{fig:biomechanics}
    \Description{ExoKit supports shoulder abduction-adduction, shoulder extension-flexion and elbow extension-flexion. According to prior work, an adult's range of motion for shoulder adduction-abduction ranges from 0 to 180 degree, for shoulder extension-flexion from -60 to 180 degree and for elbow extension-flexion from 0 to 150 degree. ExoKit supports shoulder adduction-abduction in a range from 0 to 90 degree, shoulder extension-flexion from -20 to 115 degree and elbow extension-flexion from 0 to 115 degree.}
\end{figure*}

\section{Related Work}
This work is informed by prior work on exoskeletons, robot programming, and toolkit research in HCI.

\subsection{Exoskeletons}
Robots worn on the human body are gaining increasing attention in the HCI community~\cite{inami_2022, muehlhaus_2023,saberpour_2023}.
Important types include wearable robotic limbs~\cite{saraiji_2018,yamamura_2023}, prosthetic limbs~\cite{hofmann_2016,phelan_2021}, robots that roam on the user's body~\cite{dementyev_2016,kao_2017,sathya_2022}, and exoskeletons~\cite{nishida_2020, gu_2016,li_2023}. Actuating the user's body, exoskeletons integrate particularly close with the human body. The resulting embodied partnership opens up a unique potential for designing novel, embodied human-robot interactions~\cite{li_2023}.
However, to date, research on exoskeletons in HCI is rare and how to design for intertwined human–computer integrations is not well understood~\cite{mueller_2023}. One of the few works on exoskeletons within the HCI community is Dexmo, an inexpensive and lightweight hand-based exoskeleton, used for motion capturing and force feedback in VR~\cite{gu_2016}. DigituSync, another prominent example, employs a dual hand-based exoskeleton to share hand movements between two people for skill learning through interconnected links, showcasing the potential of exoskeletons for novel and creative applications in HCI~\cite{nishida_2022}.
Beyond hand-based exoskeletons, only little research investigated the elbow~\cite{teng_2022,homola_2022}. For instance, Homola et al. explored exoskeletons to aid game-based rehabilitation, leveraging a low-fidelity elbow-based exoskeleton built from Lego~\cite{homola_2022}. 

Exoskeletons have been extensively studied in the robotics community, but contrary to the HCI community, their primary focus lies on sophisticated mechanical designs whose objectives are stability, force, the support of complex joint types, and the prevention of misalignments with the human body, along dedicated control strategies~\cite{gull_2020}.
Such exoskeletons are often developed for specific applications~\cite{gull_2020}, such as  industry~\cite{bogue_2018}, military~\cite{proud_2022}, rehabilitation~\cite{proietti_2016}, and daily life assistance~\cite{jung_2018, kapsalyamov_2020}, rather than as a generic design that allows for a broader exploration of interactions and rapid prototypability.

Contrasting the two communities, it is apparent that HCI's requirements for exoskeleton hardware inherently differ from those in robotics research. While the former focuses on user-centered design objectives, requiring rather low-fidelity prototypes that facilitate the exploration of novel human-exoskeleton interactions, the latter primarily excels in the development of sophisticated mechanical and control aspects. 
Attempting the needs of educators and tinkerers, EduExo\footnote{EduExo Kit: \url{https://www.eduexo.com/eduexo-kit/}, last accessed July 3, 2024} provides a valuable platform to teach fundamentals of exoskeleton principles to novices~\cite{tashi_2024,bartenbach_2020}.
However, it intentionally engages users with low-level motion control and provides limited actuated DoFs, and thereof still leaves a barrier to novices interested in human-exoskeleton interaction design from a higher-level perspective. Hence, there is a need for a toolkit that provides a low-fidelity exoskeleton prototype which empowers users to easily and rapidly customize the exoskeleton's design and interactive behavior to their needs, without having to handle the details of mechanical design and low-level motion control. 
We deliberately focused our toolkit on the shoulder and elbow to complement prior work in HCI which has primarily focused on hand-based and, to a lesser extent, on elbow-based exoskeletons~\cite{gu_2016,teng_2022,nishida_2022,li_2023}. 

\paragraph{Biomechanics of the upper limbs}
Considering the biomechanics of the human upper limbs is crucial for the design of arm-based exoskeletons. Key properties that are particularly relevant are the joints’ degrees of freedom (DoFs), range of motion (RoM), and torque settings. 
The shoulder joint possesses three DoFs, enabling abduction-adduction (\autoref{fig:biomechanics}a), flexion-extension (\autoref{fig:biomechanics}b), and medial-lateral rotation. The elbow has two DoFs, allowing for flexion-extension (\autoref{fig:biomechanics}c) and supination-pronation, which is the rotation of the forearm.
The RoM defines the maximum arc through which the joint can move.
Figure~\ref{fig:biomechanics} provides an overview of the DoFs and RoMs supported by ExoKit, which lay within the typical RoMs of an adult's joints~\cite{norkin_2016}.
Another critical factor in the exoskeleton design and selection of suitable actuators is the torque required for the application, as different applications may require different torques; for example, rehabilitation exoskeletons may require less torque than industrial exoskeletons designed to support heavy loads. Furthermore, a user's torque perception is also influenced by various other factors such as movement direction or speed~\cite{kim_2021}. Therefore, the required torque settings need to be carefully determined for the specific use case.

\subsection{Robot Programming}
End-user robot programming focuses on enabling non-experts to control robot operations, broadening their deployment in real-world applications. To simplify robot programming, various approaches exist, such as API-based manual programming~\cite{diprose_2017, angerer_2013}, visual block-based programming~\cite{weintrop_2018}, or programming by demonstration~\cite{ajaykumar_2021}.
One interesting API-based approach is the Robotics API~\cite{angerer_2013}, an object-oriented framework for developing complex robotic applications. It abstracts key robotic actions and sensors as object proxies, which users can sequentialize and parallelize in an event-driven architecture. Such a ``Trigger-Action Programming'' approach has been widely used in HCI, too, enabling non-experts to program devices like smart homes~\cite{ur_2014} and humanoid robots~\cite{leonardi_2019} with relative ease. 
One crucial consideration in programming robots in this way is to identify essential functional abstractions which can be combined according to the tasks the robot should perform~\cite{desantis_2008}. 
Striking a balance between exposing low-level functions for flexibility and providing high-level abstractions for ease of use is critical. While prior work has investigated this in domains like industrial or social robotics~\cite{angerer_2013,leonardi_2019}, no such abstractions have been established for exoskeletons to the best of our knowledge. We argue that identifying a suitable level of functional abstractions for novice roboticists is one crucial aspect to facilitate rapid prototyping of human-exoskeleton interactions, empowering them to focus on high-level augmentation strategies rather than low-level motion control.

\subsection{Robotics Toolkit Research in HCI}
Toolkit research plays an important role in HCI by making new technologies accessible to broader audiences, reducing complexity, encouraging creative exploration, and integrating with existing practices~\cite{ledo_2018}. HCI toolkits typically achieve these goals by encapsulating low-level domain knowledge, lowering the barrier to entry for non-experts~(e.g.,~\cite{lei_2022, arabi_2022,li_2019}). However, despite being established in HCI, toolkits specifically addressing robot prototyping remain limited~\cite{suguitan_2019}. A few notable examples include Blossom~\cite{suguitan_2019}, Phybots~\cite{kato_2012}, and WRLKit~\cite{saberpour_2023}. Suguitan et al. identify three key requirements for their robotics toolkit, two of which are particularly relevant to \toolkit: accessibility for rapid prototyping and flexibility to customize both the robot's design and behavior~\cite{suguitan_2019}. 
Similarly, Phybots is a toolkit targeted at researchers and interaction designers to rapidly prototype with locomotive robots through hardware and a software API~\cite{kato_2012}.
WRLKit enables designers to prototype personalized wearable robotic limbs tailored to specific tasks and body locations through a computational design tool, thereby lowering the entry barrier to wearable robotics for the HCI community~\cite{saberpour_2023}. 
Inspired by these approaches, we address the gap in wearable robotics toolkits in HCI with \toolkit, a do-it-yourself toolkit which lowers the entry barrier for prototyping human-exoskeleton interactions.

\section{Design Considerations}
With \toolkit, we set out to develop an exoskeleton specifically for the arms with up to 3 active degrees of freedom~(DoF) on each arm. We extend beyond existing HCI exoskeleton prototypes, which have largely focused on the hands and elbows to date (e.g.,~\cite{teng_2022,gu_2016,homola_2022, nishida_2022}). By targeting the arms including the shoulder joint, we aim to open up a wider range of applications that involve larger body movements, such as motion guidance in sports, physical rehabilitation, and experiences in VR. 
\toolkit's objective is to facilitate the prototyping of human-exoskeleton interactions for novice roboticists in the early design stages.
Its design considerations are guided by the five goals of HCI toolkit research~\cite{ledo_2018} and informed by literature on wearables~\cite{gemperle_1998, saberpour_2023} and exoskeletons~\cite{souza_2016,sarac_2019}:

\paragraphB{Ease of development}
Programming the interactive behavior of robots is a challenging task for novices~\cite{ajaykumar_2021}. To ease development and empower designers to focus early on on the prototyping of novel interactions, \toolkit~should
\begin{itemize}[leftmargin=*, noitemsep, topsep=3pt]
\item provide \textit{powerful functional abstractions} that offer relevant functionalities for programming versatile interactive behaviors in few lines of code. In line with principles of end-user robot programming (e.g.,~\cite{desantis_2008}), these should abstract away the details of low-level motion control and instead allow the developer to focus on higher-level augmentation strategies,  
\item enable \textit{interactive reactions from sensing to actuation}. As a body-centred technology, it is essential to provide functionalities that sense and interpret body motions in real-time, such as a limb's motion direction, position, or speed. This allows designers to employ these as conditional triggers for specific exoskeleton actuation, creating complex interactive behaviors.
\item should provide an infrastructure that allows to \textit{experience the implemented interactions with the prototype in real-time}, thus enabling designers to iteratively improve the design -- a key principle in interaction design~\cite{preece_2015}.
\end{itemize}

\paragraphB{Fabrication}
In line with design principles discussed in other physical toolkit papers in HCI, \toolkit~should
\begin{itemize}[leftmargin=*, noitemsep, topsep=3pt]
\item use \textit{accessible materials and fabrication processes}. To empower new audiences and  integrate with existing practices~\cite{ledo_2018}, we intend to use technologies widely used in HCI, design and making~(e.g., 3D printing, Arduino) for building and implementing the prototype. This allows the community to use, extend and tailor the toolkit to their needs.
\item consist of \textit{modular components} that facilitate easy assembly of the prototype and rapid exchange of functional components after fabrication. This aligns with the iterative nature of interaction design and prior approaches of robot toolkits in HCI~\cite{cui_2023,cui_2024}.
\item be \textit{customizable to individual body characteristics}, an essential consideration in wearable robotic toolkits~\cite{saberpour_2023}. As different body characteristics can affect an exoskeleton's fit, function and wearability, \toolkit~must offer mechanisms to easily adjust the prototype dimensions to different arm lengths and circumferences. 
\end{itemize}

\paragraphB{Safety}
\toolkit~must involve \textit{safety measures}, which are essential for physical HRI~\cite{desantis_2008, saberpour_2023}. Actuating the human body, we must provide mechanisms that ensure that the applied forces and ranges of motions are safe and comfortable~\cite{souza_2016}, even in case of inexperienced developers.  This also involves, for instance, to accommodate motors with different maximum torques to let designers decide what is a safe range required for their application. 
\begin{figure}[b]
    \includegraphics[width=0.45\textwidth]{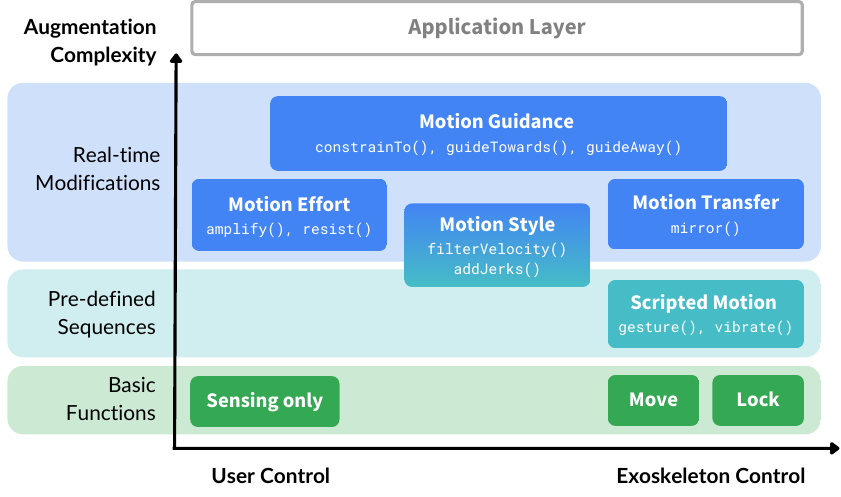}
    \caption{~\toolkit's key functionalities comprise basic functions and five augmentation strategies. They are conceptually organized in a two-dimensional space: Control determines whether the user is in full control of movement, is guided or constrained, or fully controlled by the exoskeleton. Augmentation complexity captures the degree of abstraction, from basic functions up to real-time modification of user movement.}
    \label{fig:conceptual_space}
    \Description{Two-dimensional space of ExoKit’s interaction concepts, with the x-axis ranging from user control to exoskeleton control, and the y-axis showing increasing augmentation complexity. The lowest complexity layer includes 3 basic functions: "Sensing only”,” Move" and "Lock". The middle layer comprises pre-defined sequences which feature the “Scripted Motions” strategy. The highest complexity is given by the real-time modifications layer, which covers "Motion Effort”, "Motion Style”, "Motion Transfer”, and "Motion Guidance”. These are described in more detail in the text.}
\end{figure}
\begin{figure*}[h]
    \centering
    \includegraphics[width=0.77\linewidth]{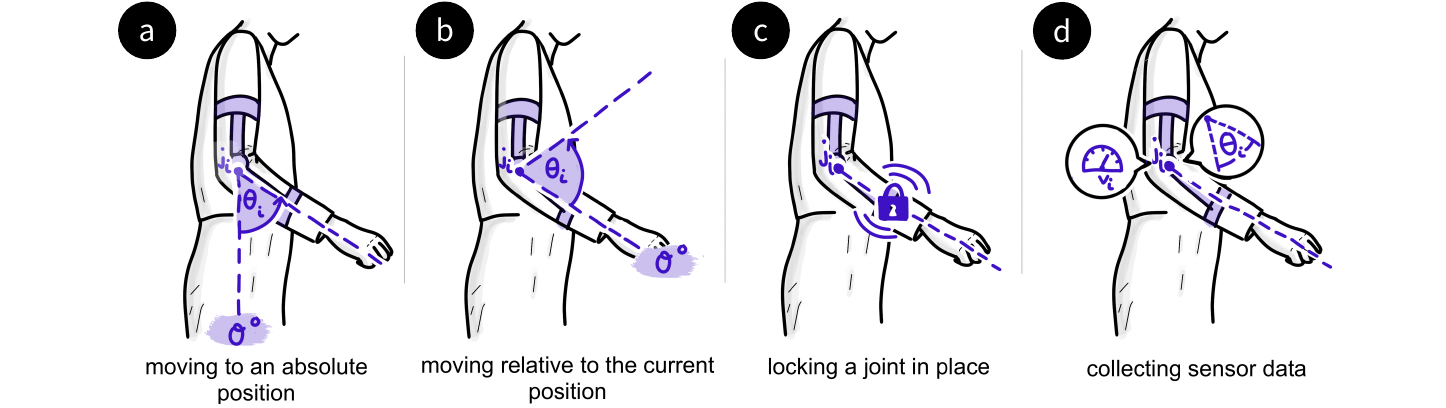}
    \caption{Basic functions offered by \toolkit: Moving a joint $j_i$ to an angle $\theta_i$ expressed as (a)~absolute w.r.t. $j_i$'s calibrated zero-degree angle, or (b)~relative to the joint's current angle; (c)~locking joint $j_i$ in place, or (d)~collecting real-time sensor data of selected joints.}
    \label{fig:move_to}
    \Description{The figure illustrates the basic functions of ExoKit as detailed in the caption and accompanying subsection.}
\end{figure*} 
\section{\toolkit's Interaction Concepts}\label{sec:concepts}

\toolkit's objective is to facilitate the design of interactions for arm-based exoskeletons. For this purpose, it provides functional abstractions, which designers can combine into meaningful interactions. 
Beyond basic sensing and actuation, we identified five essential clusters of motion augmentation strategies that are recurrently used in the state-of-the-art exoskeleton literature~\cite{gasperina_2021,proietti_2016,gull_2020,fernandez_2023,basteris_2014}. We have organised them in a conceptual space, which is defined by two key dimensions of motion augmentation~(see Figure~\ref{fig:conceptual_space}): \textit{control} and augmentation \textit{complexity}. 
\textit{Control} captures the extent of control the user vs.~the exoskeleton have over the movement. At one end of the spectrum, the user is in full control and therefore can move freely to any target position. As we move along this axis, the user's freedom is progressively reduced while the exoskeleton is taking over more control, up to a point where the exoskeleton fully controls the user's motion.
Augmentation \textit{complexity} starts from basic functions, like sensing and moving, to pre-defined sequences of basic functions, up to augmentation strategies that modify a user's motion in real-time through a feedback loop.

In the following, we start with presenting the basic functions, followed by the higher-level motion augmentation strategies: 

\subsection{Basic Functions}
Exoskeletons built with \toolkit~are composed of either one or two arms. Each arm consists of up to three joints $j_i, i \in \{1,2,3\}$, which each can be either passive, sensing, or actuated. 
As a prerequisite for the following actuation functionalities, the designer must register the exoskeleton's configuration and calibrate the absolute zero-degree position in software for each actuated or sensing joint $j_i$ once at system startup.

\paragraphB{Actuation}
An actuated joint $j_i$ can be moved to a certain position. As body-centered technology, exoskeleton motions can be defined through angles $\theta_i$ in the user's joint angle space. 
Here, we distinguish between angles that are absolute with respect to the calibrated zero-degree angle of joint $j_i$, and relative angles which describe an angle relative to the user's current position~(cf., Figure~\ref{fig:move_to}a, b).
\toolkit~enables designers to move a joint $j_i$~(\textit{moveTo()}) by specifying an absolute or relative target angle $\theta_i$, the range $\epsilon$ around $\theta_i$~(useful for approaching an area centered at $\theta_i$) and the velocity $\upsilon$. 
To fully prevent any user motion, a joint $j_i$ can also be locked (\textit{lock()}). This is achieved by activating the torque of the motor. Internally, the motor then maintains the position by continuously applying counterforces to the user's motion, thereby providing a locking experience for the user~(Figure~\ref{fig:move_to}c). 
For ease of programming, all functions are provided for controlling an individual joint only, a set of joints, or the entire arm at once. 

\paragraphB{Sensing}
\toolkit~provides real-time sensor data about the exoskeleton's physical angle configuration, motion velocity, acceleration, and applied torques~(Figure~\ref{fig:move_to}d). 
\toolkit~can continuously stream this sensor data from selected or all joints through a serial port connection, making the data easily accessible to external applications. Secondly, \toolkit~allows to use sensor data as conditional triggers in the firmware library: Developers can configure basic threshold- and range-based conditions about the exoskeleton's physical configuration~(speed, acceleration, angle and torque). For instance, these help to detecting when the user moves outside a specified range of motion or exceeds a certain speed. Further, designers can access more complex movement states, which provide insights into each joint's motion direction and help to assess if the user's arm is at a desired angle configuration (or \textit{pose}). 

\begin{figure*}[t]
    \centering
    \includegraphics[width=\linewidth]{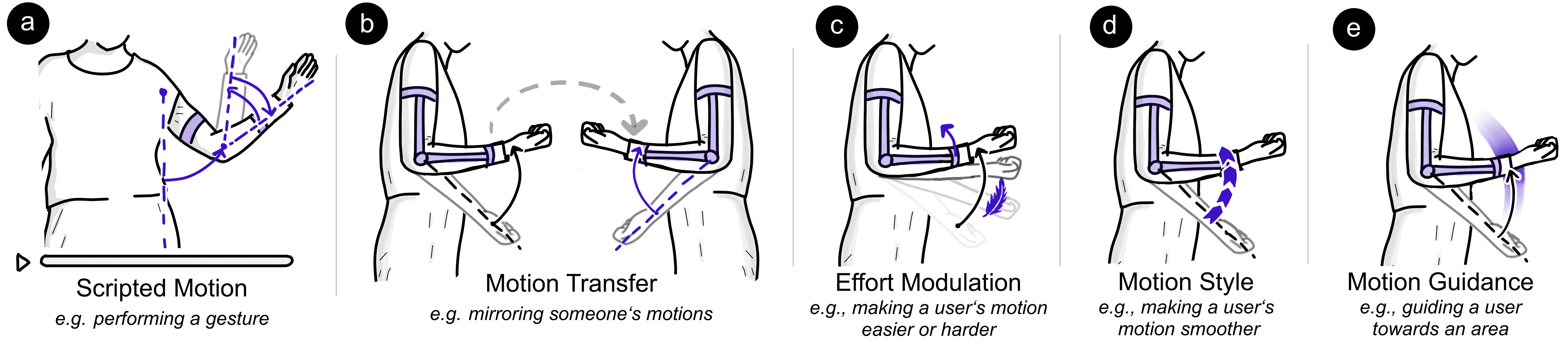}
    \caption{\toolkit~provides functional abstractions for five relevant classes of augmentation strategies which enable to (a)~replay a scripted motion on demand, (b)~continuously transfer one user's motions onto another one, (c)~augment a user's motion effort by amplifying or resisting her ongoing motions, (d)~alter the user's motion style, or (e)~guide the user towards or away from an area.}
    \label{fig:interaction_concepts}
    \Description{The figure illustrates functional abstractions provided by ExoKit as detailed in the caption and accompanying subsections.}
\end{figure*}

\subsection{Scripted Motions}
Building on the basic functions, designers can create and playback scripted sequences of motion, which are defined as a time series of basic move commands. 
Scripted body motions can be helpful to realize a specific repetitive motion with fixed parameters, or to play back motion sequences that serve as gestures~(e.g., waving) and kinesthetic notifications~\cite{catanua_2023}.
For illustrative purposes, we provide functions for executing a waving gesture with the arm~(\textit{gesture()})~(see Figure~\ref{fig:interaction_concepts}a), and for generating a vibrating sensation through rapid back-and-forth movements of adjustable amplitude and frequency~(\textit{vibrate()}). The latter serves as a haptic notification for the user which can be delivered either to a single joint, a set of joints or the full arm.

\subsection{Motion Transfer}

The concept of motion transfer involves transferring human motion in real-time from one joint to another joint, from one arm to another arm, or even from one user to another person. This offers flexible means for applications: These involve guiding another user's motion, such as for motion teaching and skill transfer ~\cite{nishida_2022,maekawa_2019} but also for creating shared kinesthetic experiences between users, or for programmatically remapping body input to output, such as mirroring an arm's movement to the other arm~\cite{pignolo_2012} or leveraging unemployed degrees of freedom for bodily control. 
\toolkit~offers functions that facilitate motion transfer between two joints or exoskeleton instances, where in every time interval, the state of the sensing joints is read and the mapped actuated joints are moved accordingly. 
Designers can mirror the input signal from a sensing joint onto an actuated counterpart~(\textit{mirror()})~(see Figure~\ref{fig:interaction_concepts}b) and optionally scale the motion up or down by a user-defined factor~$f$, while the system internally ensures that the scaled motion does not exceed the user's calibrated limits of their range of motion. 

\subsection{Modulating the Motion Effort}
Modulating the motion effort comprises interaction strategies in which the exoskeleton applies torques $\tau$ in or opposite to a user's inherent motion, while preserving the user's ability to freely navigate in space. 
The resulting feeling of physical assistance or resistance can be leveraged by designers to modulate how users perceive their own motion, with stronger torques $\tau$ resulting in a stronger effect. 
Examples where effort modulation can prove beneficial are rehabilitation, where assistance eases a patient's movements while resistance enhances the muscle strengthening~\cite{gasperina_2021,proietti_2016}, or virtual reality, where the immersiveness can be enhanced by adjusting the level of effort to better match the virtual environment~\cite{teng_2022}.
\toolkit~provides two pre-implemented functions which designers can use to increase~(\textit{resist()}) or decrease~(\textit{amplify()}) the motion effort, respectively, selectively for individual joints or applying to the entire exoskeleton. These functions include parameters that allow designers to adjust the assistive or resistive torque $\tau$ and to specify whether the strategy should be continuously active or only triggered when the user moves in a certain direction~(see Figure~\ref{fig:interaction_concepts}c).

\subsection{Modulating the Motion Style}

Modulating the motion style refers to interaction strategies that modify the observable characteristics of a user’s movement, such as speed or path, while keeping the user in control of the overall motion. These style modifications can range from subtle adjustments to more pronounced changes.
Examples of motion style modulation from the literature include exoskeletons that suppress tremor~\cite{lora_2021}, or exoskeletons that add perturbations to a user's movement, challenging their motion control~\cite{gasperina_2021}.
\toolkit~provides two pre-implemented functions to facilitate exploring modulations of motion style: (1)~The exoskeleton tries to keep a user's motion speed within a pre-defined range~(\textit{filterVelocity()}), by applying assistive torques $\tau$ if they are too slow or resisting ones if they are moving too fast. Designers can fine-tune these effects by setting the velocity range and the forces applied to maintain it. (2)~As an example of an artistic style modulation, the exoskeleton can introduce jerks~(\textit{addJerks()}) to augment a user's motion path for a more technical or robot-like motion style~(see Figure~\ref{fig:interaction_concepts}d). A jerk consists of a short randomized movement that creates a distortion in the user's movement. Designers can adjust this effect by specifying the minimum and maximum angular displacements, the time intervals between jerks, and the number of jerks.

\subsection{Motion Guidance}
\begin{figure}[bt]
    \centering
    \includegraphics[width=\linewidth]{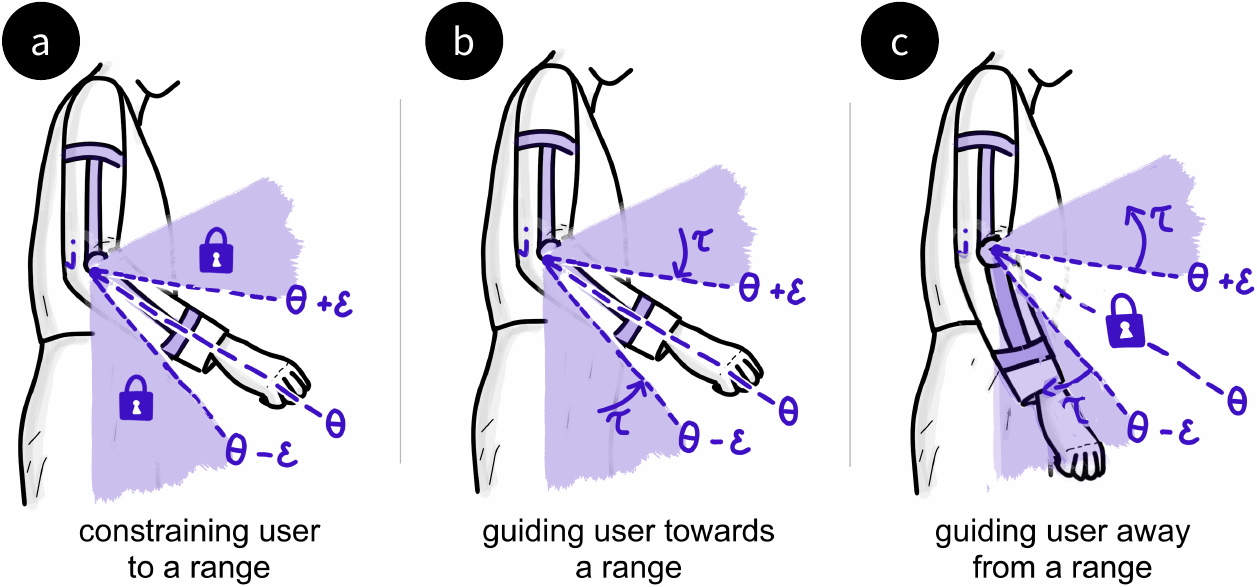}
    \caption{\toolkit's software features three functions for motion guidance: (a)~constraining a user's motion to an area centered around $\theta_i$ with range $\epsilon$, (b)~guiding a user towards this area, (c)~or guiding the user away from it.}
    \label{fig:guidance}
    \Description{The figure illustrates the motion guidance strategies implemented in ExoKit, as detailed in the caption and accompanying subsection.}
\end{figure}

Haptic guidance comprises strategies in which the exoskeleton applies forces with the goal of guiding a user towards a certain point or area. These strategies are situated in-between full user control and full exoskeleton, where both effectively collaborate toward a shared goal. Depending on how parameters are adjusted, the designer can trade-off user control vs.~exoskeleton control: exoskeleton control increases as guiding forces increase and as the range of motion becomes more constrained.
Examples of haptic guidance include tunneling approaches in rehabilitation, where the patient’s movement is locked within a predefined path or tunnel, and force fields that guide them back toward the center of it~\cite{guidali_2011,proietti_2016,gasperina_2021}. Beyond rehabilitation, they also hold promise in other areas, such as gaming and VR. 
\toolkit~provides three complementary functions~(see Figure~\ref{fig:guidance}):
(1)~The exoskeleton can constrain the range of motion of a joint $j_i$ to an area around an absolute angle $\theta_{i}$ with range $\epsilon$~(\textit{constrainTo()}). As soon as the user attempts to move beyond this range, the exoskeleton tries to move the user back to the boundary with maximum torque, thereby keeping the joint within the specified limits. 
(2)~The exoskeleton can guide the user towards an area centered around $\theta_i$ with range $\epsilon$~(\textit{guideTowards()}). Here, the exoskeleton applies assistive torques $\tau$ to $j_i$ when moving towards the desired area and resists otherwise~(see Figure. (3)~Lastly, the exoskeleton can keep a joint $j_i$ away from the area around $\theta_{i}$~(\textit{guideAway()}), applying resistance with torque $\tau$ if the user is approaching the area and assistive torques otherwise. In addition to defining angle $\theta_{i}$ and range $\epsilon$, designers can also specify the magnitude of the amplifying and resisting forces.

\subsection{Programming Interfaces}\label{subsec:interfaces}

\begin{figure}[bt]
    \centering
    \includegraphics[width=\linewidth]{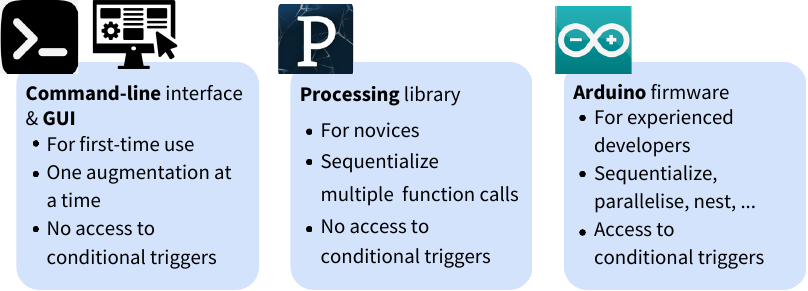}
    \caption{\toolkit~provides a command-line interface and GUI, a Processing library, and an Arduino firmware library to program human-exoskeleton interactions.}
    \label{fig:programmability}
    \Description{Overview of ExoKit’s programming interfaces and their key features. The command-line interface and graphical user interface are designed for first-time users, allowing to call one augmentation at a time without access to conditional triggers. The Processing library targets novices which can define sequential function calls. The Arduino firmware is intended for experienced developers, supporting sequential, parallel, and nested operations with access to conditional triggers.}
\end{figure}

\toolkit~provides different interfaces to program the interactions, illustrated in Figure~\ref{fig:programmability}. Each interface is targeted at different experience levels, a recommended practice for end-user robot programming~\cite{ajaykumar_2021}. 
First, for novices in the early exploration stages, a simple command-line interface~(CLI) can be used for executing the basic functions and higher-level augmentation strategies. It abstracts away complex programming concepts such as conditions or loops. We also offer a graphical user interface~(GUI) programmed with Processing, which can be used in place of the CLI. The GUI allows novices to execute basic functions at the touch of a button, and tune parameters with sliders. Second, users can leverage a Processing library to define sequences of functions and strategies, enabling the programming of slightly more complex interactive behaviors.
Third, more experienced users can program complex behavior within the Arduino firmware. The firmware builds on trigger-action programming, realized through an object-oriented event-driven architecture, in which users can combine basic functions and higher-level augmentation strategies into their own customized interactions. Similar to other works on end-user robot programming ~(e.g.,~\cite{guerin_2015, angerer_2013, ur_2014}), users can define elements of parallelization, sequences, nested functions, or leverage incoming sensor data as conditional triggers. For more implementation details, see Appendix~\ref{subsec:implementation_sw} and the supplemental material. 

\begin{figure*}[t]
    \centering
    \includegraphics[width=0.8\linewidth]{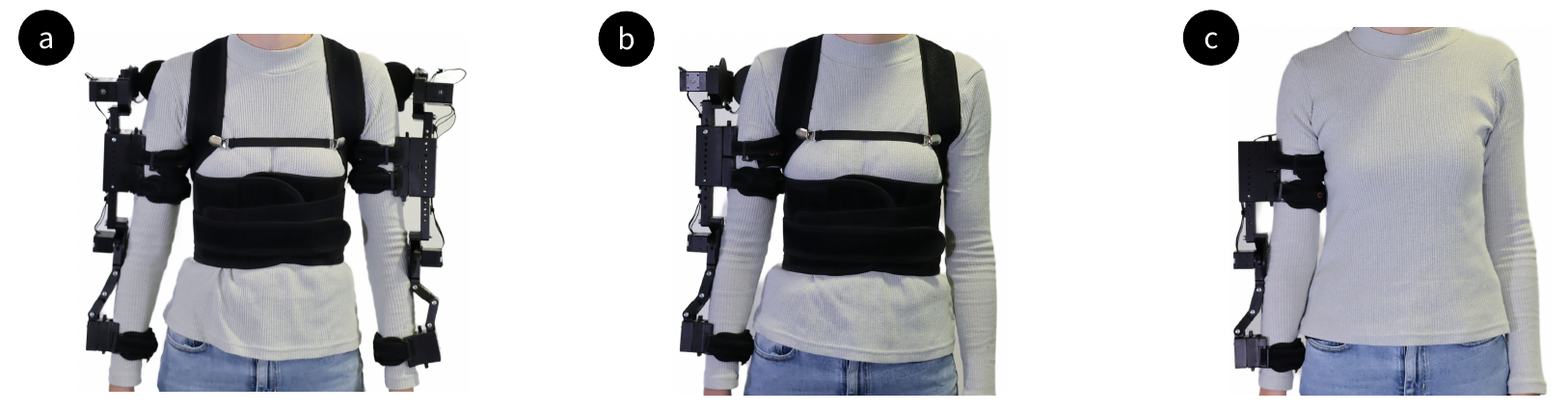}
    \caption{\toolkit's modular components allow to freely configure the degrees-of-freedom (DoF). Some example configurations include: (a)~6 DoFs at both arms, (b)~3 DoFs at one arm, (c)~1 DoF at one elbow.}
    \label{fig:customizable_dof}
    \Description{Examples of ExoKit’s adjustable degrees-of-freedom. The image features three sub-images of a person wearin ExoKit configured with varying degrees-of-freedom. The caption provides additional details. In configurations that involve the shoulder, the person also wears a posture corrector, which is used to secure the exoskeleton to the body.}
\end{figure*}
\section{\toolkit~Hardware Components}\label{sec:hardware}
A key challenge in designing hardware components for an exoskeleton toolkit lies in balancing functional requirements, such as required degrees of freedom, stability, and safety, with designer's needs for rapid prototyping. Particularly in the early stages of interaction design, an important key characteristic for design is iteration~\cite{preece_2015}, in which a designer usually explores a wide range of possibilities before converging on the most promising solutions. 
While modularity, hands-on reconfigurability, and prototypability are well-established principles in HCI robotic toolkits (e.g.,~\cite{cui_2023,cui_2024}), these considerations are less prevalent in exoskeleton research so far. Traditionally, research on exoskeletons prioritizes functional requirements for high-fidelity solutions for specific applications, driven by the unique demands of each use case (e.g.,~\cite{sarac_2019}). 
We therefore note a disciplinary gap between the iterative and exploratory prototyping approach required in HCI and the function-focused demands of professional, specialised exoskeletons.

Building on the design considerations from Section 3, the design of \toolkit~'s hardware components is centered on three aspects that integrate exoskeleton-specific requirements~\cite{souza_2016, sarac_2019} with HCI principles: (1)~Modular components that facilitate hands-on reconfigurability of the exoskeleton's functionality, involving their DoFs, sensing and actuation capabilities, (2)~alignment strategies that facilitate adjusting the exoskeleton's fitting for wearability and rapid prototyping, and (3)~safety mechanisms suited for inexperienced developers.

All components offered by \toolkit~can be fabricated using commercially available 3D printers and ABS filament, addressing the trade-off between stability and accessible fabrication. The toolkit further relies on widely-used off-the-shelf components, notably Arduino and Dynamixel motors.
The current \toolkit~prototype supports all motors of the Dynamixel XM series, providing forces up to $10~Nm$ with the strongest motor, and a maximum speed of $77$ RPM with the fastest. The motors can be powered from 12 Volts using the supplied Dynamixel power supply or a wearable power bank. 
For torque sensing, the toolkit achieves rates of $80~Hz$. 
\toolkit~allows to configure exoskeletons with up to 6 active degrees of freedom -- 3 per arm. The supported ranges of motion are $0$ to $115$ degrees for elbow flexion-extension, $-20$ to $115$ degrees for shoulder flexion-extension, and $0$ to $90$ degrees for shoulder abduction-adduction, comparable to existing exoskeleton designs (cf.,~\cite{rahman_2015,bilancia_2021,kiguchi_2008}). More implementation details can be found in Appendix~\ref{subsec:implementation_hw}.

In the following, we detail on the individual hardware components offered in \toolkit~and their roles in enabling reconfigurable, size-adjustable and safe exoskeletons:

\subsection{Modular Components To (Re)Configure the Exoskeleton's Functionality}\label{subsec:reconfigurability}

In early stages of prototyping, designers might not yet have a clear idea of the exoskeleton's final functional requirements. Thus, it is essential to create exoskeletons in HCI that support iterative development by empowering designers to add, remove, and modify components as requirements and the understanding of the problem space evolve.
To address this, we propose a principled, modular architecture that is based on three key considerations:
(1)~In-line with modular exoskeleton designs~\cite{souza_2016}, the exoskeleton should allow designers to adjust the DoFs to suit different application scenarios. (2)~As exoskeletons serve a range of purposes---from motion capture~\cite{gu_2016} to delivering haptic feedback~\cite{teng_2022}, or applying stronger actuation forces~\cite{exoJacketIndustrial}---designers must be able to swap between sensing and actuating modules of varying strengths to match specific task requirements. (3)~Inspired by exoskeletons where sensors extend beyond position sensing in joints (e.g., load cells~\cite{gull_2020}), the architecture must be able to accommodate additional sensing elements, ensuring that the exoskeleton can be adapted to changing requirements. We now discuss each consideration:

\paragraphB{Customizable degrees-of-freedom}
Different applications require varying DoFs in exoskeletons.
The challenge for an exoskeleton toolkit with customizable DoFs lies in providing the necessary flexibility by freely configuring the DoFs after fabrication, without adding mechanical complexity, more components or reducing user comfort. Moreover, the provided components, even for more complex joints such as the shoulders, should be fabricable using accessible fabrication methods while providing a sufficient degree of stability.
The toolkit extends beyond current exoskeleton approaches in HCI by offering support for the elbow and the more complex shoulder joints. It supports shoulder flexion/extension and abduction/adduction through two DoFs which can be actuated and a passive DoF that enables the exoskeleton to adapt to complex movements of the shoulder joint. 
In addition, the elbow is supported with one DoF in flexion-extension. 
The designer can easily adjust the hardware prototype by detaching joints to remove DoFs not needed for a specific application. For instance, designers can construct exoskeletons that support both arms, only one arm or the elbow (see Figure~\ref{fig:customizable_dof}).

\begin{figure*}[tb]
    \centering
    \includegraphics[width=1\linewidth]{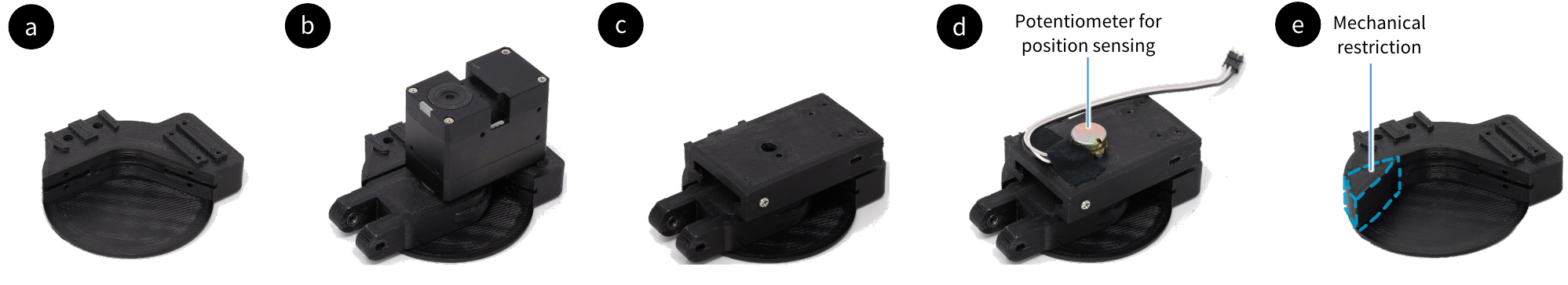}
    \caption{\toolkit~features interchangeable joints based on a versatile joint base (a). This base can be fitted with (b)~a motor for actuated joints, (c)~a passive module to replace the motor, or (d)~a passive module with an integrated position encoder for sensing-only joints. (e)~Designers can insert mechanical restrictions to the base joint to further limit the offered range of motion by 15, 30 or 45° on each side.}
    \label{fig:joint_modules}
    \Description{The figure illustrates the different joint types employed in ExoKit, as detailed in the caption and accompanying subsection.}
\end{figure*}

\paragraphB{Modular joints}
As exoskeletons can be either actuated, act as a sensing device, or be passive, it is crucial for designers to have the flexibility to choose the functionality for their application themselves. To enable hands-on reconfigurability, \toolkit~offers three types of interchangeable joint modules: (1)~\textit{Actuated joints} comprise an actuator. A joint base, depicted in Figure~\ref{fig:joint_modules}a, serves for mounting a motor (Figure~\ref{fig:joint_modules}b). The base is designed to be compatible with all motors of the Dynamixel XM series. These motors inherently offer sensing capabilities (position, velocity). If desired, developers can modify the design of the joint base to include other motors with different dimensions. 
(2)~\textit{Sensing-only joints} measure the exoskeleton angle configuration, through which further motion properties, including velocity and acceleration, can be inferred. (3)~\textit{Passive joints} simply follow the user's motions, realizing a passive DoF. 
To realize sensing-only and passive joints, \toolkit~provides a module that can be mounted on top of the joint base instead of a motor~(Figure~\ref{fig:joint_modules}c). 
This module offers a space for an optional position encoder, hence can be repurposed as a sensing-only joint when the position encoder is attached~(Figure~\ref{fig:joint_modules}d). 

\begin{figure}[tb]
    \centering
    \includegraphics[width=0.96 \linewidth]{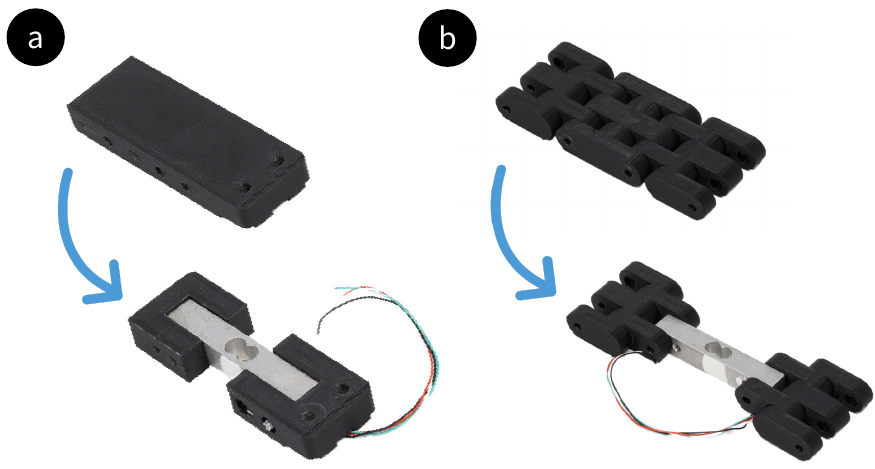}
    \caption{(a)~Rigid and (b)~flexible links without and with integrated load cells, respectively.}
    \label{fig:loadcell}
    \Description{The figure illustrates the rigid and flexible links once without and once with a load cell integrated, as detailed in the caption and accompanying subsections.}
\end{figure}

\paragraphB{Modular links}
Exoskeleton links are structural components that connect and transmit forces between joints. In \toolkit, we provide two types of 3D-printable passive links. The first is a simple rigid rod, used to stably connect the shoulder and elbow joints. The second is a flexible link made of 3d printed chain segments informed by the mechanism presented by Tiseni et al.~\cite{tiseni_2019}. The advantage of this approach is that the mechanism remains rigid in the direction where torque is applied while complying with user's complex body movements in other directions. In \toolkit, we employ this mechanism as a flexible link between the two shoulder joints, which helps to better comply with the shoulder's additional degrees of freedom, and as a link that connects the elbow joint with the lower arm cuff. 
If desired, both the rigid and chain links can be equipped with an additional load cell that captures the torques applied to the joints. As this is a commonly used control input in exoskeletons~\cite{gull_2020}, this empowers designers to quickly integrate the load cell into their design without having to manually redesign the hardware. The Arduino firmware provides complementary support, allowing users to register the load cell as an exoskeleton component, use the sensing function to stream load cell measures to external applications and deploy them as runtime triggers in the event-based architecture. 
The link modules are depicted in Figure~\ref{fig:loadcell}a and~\ref{fig:loadcell}b.

\subsection{Modular Components to Support Size-Adjustability}
\begin{figure}[tb]
    \centering
    \includegraphics[width=0.92\linewidth]{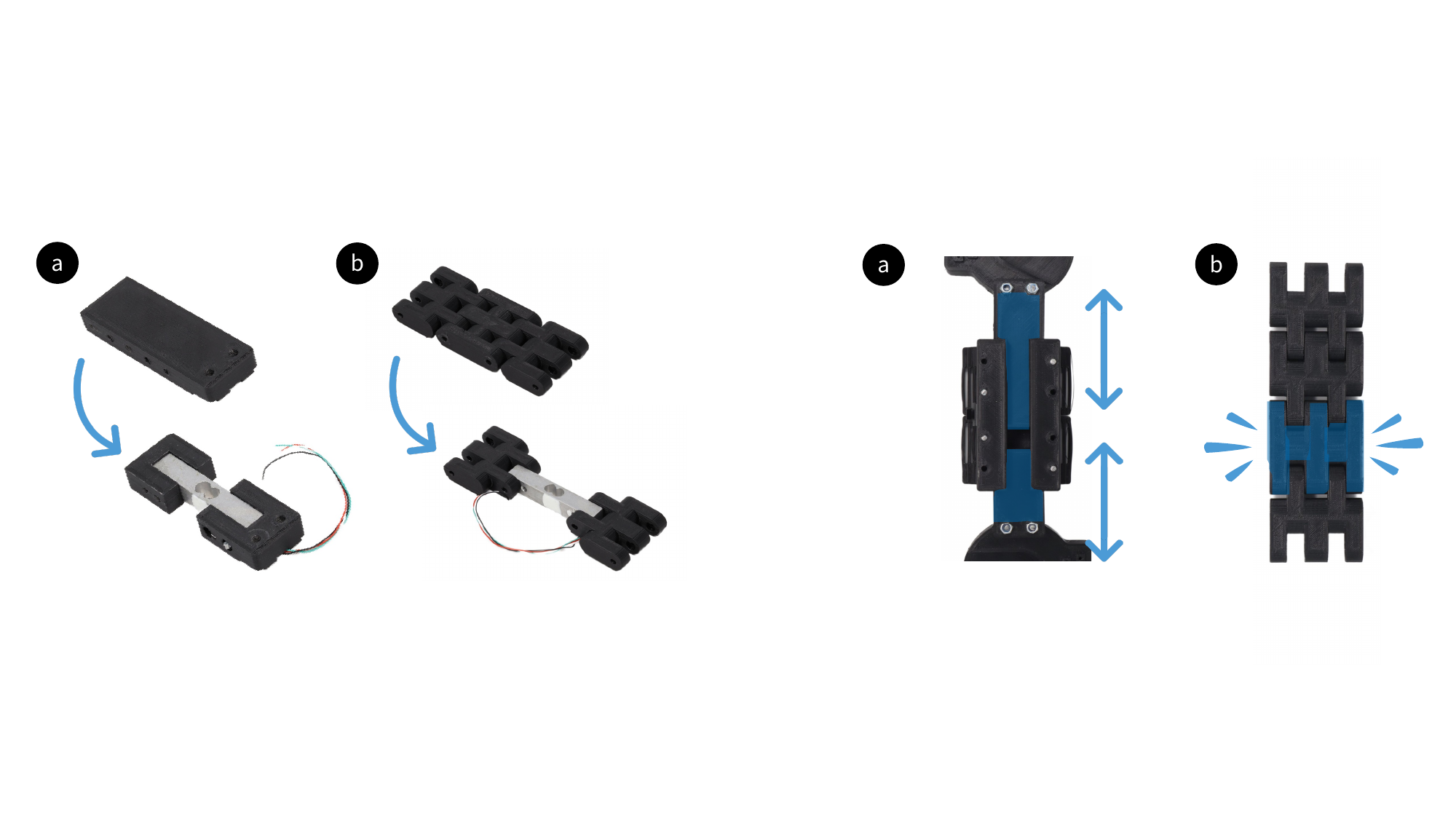}
    \caption{(a)~The rigid joint can be adjusted in length through a rail system which allows to move the 3d printed rods further in and out. (b)~The flexible link can be easily adjusted in size by adding or removing chain segments.}
    \label{fig:size-adjustability}
    \Description{The figure illustrates the size-adjustability of the rigid and flexible links, as detailed in the caption and accompanying subsections.}
\end{figure}

Adjusting the exoskeleton to the user's body size is essential to prevent misalignments that could cause discomfort or hinder proper functionality through undesired torques~\cite{gull_2020}. There are two promising approaches to achieve better joint alignments: fabricating a personalized exoskeleton or integrating size-adjustable mechanisms post-fabrication~\cite{sarac_2019}. \toolkit~follows the latter, enabling hands-on customization which avoid the need to reprint new parts for every user, thereby facilitating rapid prototyping in collaborative settings. All of \toolkit's links are designed for size adjustability. In line with strategies presented in exoskeleton literature~\cite{bartenbach_2016,zhang_2023}, the rigid links on the upper arm and back use adjustable rail mechanisms to accommodate a range of body sizes within the 95th percentile~\cite{gordon_1988}~(see Figure~\ref{fig:size-adjustability}a). The chain link introduced in Subsection~\ref{subsec:reconfigurability} can be adjusted to different shoulder circumferences and lower arm lengths by simply attaching or detaching chain segments~(see Figure~\ref{fig:size-adjustability}b). Finally, the arm cuffs which connect the links to the human arm come in two different sizes, and can be stacked together to provide a larger area that holds the exoskeleton in place.

\subsection{Physical Safety}
Safety is an important concern in exoskeletons, particularly if large forces are involved. 
Based on considerations in robotics literature~(e.g.,~\cite{souza_2016,zacharaki_2020}), \toolkit~comprises three layers of safety, which cover mechanical, software, and user-centered safety measures, specifically targeted at novice developers:

The first layer presents mechanical safety measures to maintain a safe range of motion, an important safety goal in exoskeletons~\cite{souza_2016}. Designers can add 3d printed mechanical restrictions to the base joints~(see Figure~\ref{fig:joint_modules}e). Through a simple plug-and-play mechanism, these mechanical restrictions physically prevent the motor from exceeding the defined range of motion. They are offered in 3 different sizes~(restricting by 15, 30 or 45° on both sides) and can be added, removed, or exchanged after fabrication as needed. 

The second layer realizes safety mechanisms in software, another essential building block of user safety in exoskeletons~\cite{souza_2016}. \toolkit~implements a routine continuously monitoring whether the user's configured range of motion is exceeded and if so terminates and shuts down the system. 

The third layer comprises user-initiated safety measures, to empower the user to actively stop the exoskeleton if required. \toolkit~incorporates a panic button, a common feature in human-robot interaction~\cite{zacharaki_2020}, which the user can either hold in the hands or attach somewhere on their body. Analogously to the prior layer, if triggered, the system immediately terminates the program and shuts down. This can be a particularly helpful measure to ensure safety during initial development or debugging of some new functionality. 

\begin{figure*}[t]
    \centering
    \includegraphics[width=\linewidth]{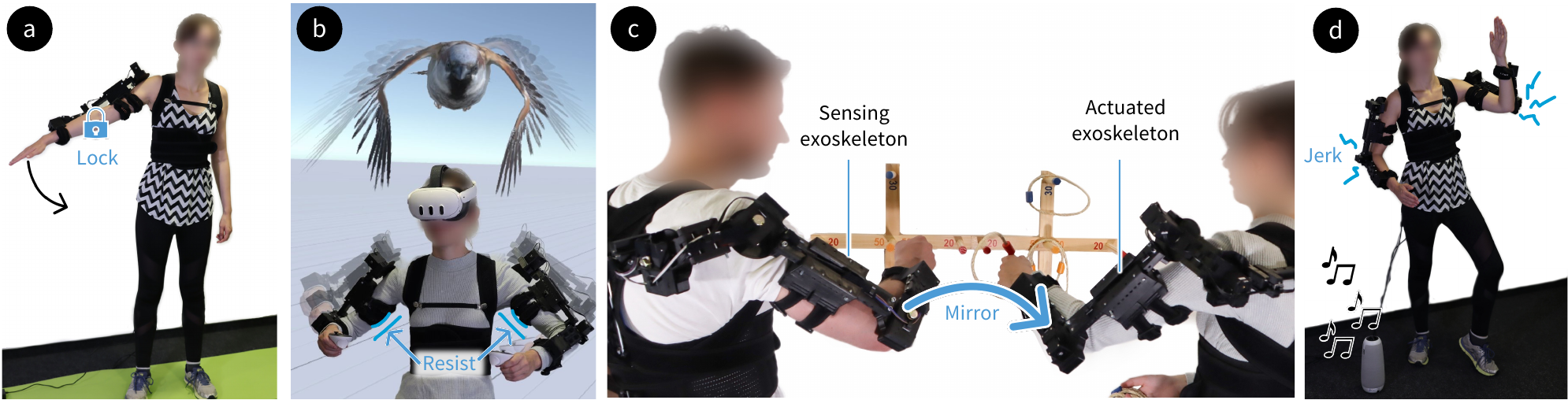}
    \caption{\toolkit~can be used for versatile applications, such as for (a) physical motion guidance for strength exercises, (b) haptic feedback in VR, (c) collaborative body-actuated plays, and (d) artistic modulation of motion in real-time.
    }
    \label{fig:applications}
    \Description{ExoKit applications. (a) A person wears the exoskeleton with three actuated DoFs at the right arm. The person performs a strengthening exercise for the arms. The application is described in detail in the corresponding paragraph. (b) A user wears a 6 DoF exoskeleton and a VR headset. The user controls a virtual bird’s wings in VR while the exoskeleton resists the user’s motion to mimic air resistance. (c) Two people are playing a game. One person wears a sensing exoskeleton arm and the other a fully actuated one. The first person controls the other's arm movements to place colored rings on a repurposed ring toss game scaffold. (d) A person is dancing to music, while the exoskeleton adds jerks to arm movements through six actuated DoFs.}
\end{figure*}
\section{Validation}
We follow Ledo et al.'s strategies for evaluating HCI toolkits~\cite{ledo_2018}: (1)~With implemented functional example applications, we demonstrate how \toolkit~enables designers to easily customize exoskeleton's design and interactive behavior. (2)~With our first usage study, which combines elements of walkthrough demonstrations and free exploration, we gather insights in \toolkit's utility. (3) Finally, with a second usage study, in which participants built their own application with ExoKit, we gained insights into users' workflows and encountered challenges.

\subsection{Example Applications}\label{sec:applications}
We illustrate the broad applicability of \toolkit~ with four implemented application examples. These are illustrated in Figure~\ref{fig:applications}.
The examples have been selected to cover different application areas, exoskeleton configurations, and augmentation strategies. All have been implemented with the Arduino firmware. We report on the first one in more detail to demonstrate the workflow. 

\paragraphB{Designing motion guidance for strength exercises}
A promising area of exoskeletons is rehabilitation and home therapy. We used \toolkit~to assist in the correct execution of an upper arm exercise aimed at strengthening muscles, inspired by an example provided in Physio@Home~\cite{tang_2015}. In this exercise, the user stretches their arm out sideways and moves it up and down repetitively, ensuring the arms don’t leave the body plane. At the same time, the elbow must remain fully extended throughout the exercise. The exoskeleton’s role is to help the user maintain proper form during execution (Figure~\ref{fig:applications}a).
We began by addressing how to keep the user's shoulder movement aligned within the body’s side plane. To do this, we built an exoskeleton with an actuated joint at the shoulder (on the side) to modify sideway shoulder motions, along with a passive joint at the back. Initially, we used the \textit{constrainTo} function to lock the user's movement within a narrow area inside the desired body plane. However, we noticed that this reduces the user’s agency substantially. To give the user more control, we switched to the \textit{guideTowards} function, which guides the user back toward the desired plane. Since the shoulder can exert significant forces during this motion, we leveraged a stronger Dynamixel motor (XM540-W270-T). This motor provides enough power to apply both assisting and resisting forces at the shoulder and we fine-tuned the torques to be effective yet gentle. 
Once satisfied with the shoulder motion, we integrated an actuated joint for the elbow, using the weakest Dynamixel motor (XM430-W210-T) and added a \textit{lock} function to ensure that the elbow does not drift from the desired position as the user moves.

\paragraphB{Kinesthetic feedback for avatar embodiment in VR\@}
An important research area in VR is providing haptic feedback~\cite{teng_2022,gu_2016}. We implemented a VR environment that uses \toolkit~to create immersive kinesthetic feedback for embodying the motion of an avatar. In our application, a flying game, the user can control the motion of an avatar's wings, by moving one's arms, and feel the corresponding kinesthetic real-time feedback~(see Figure~\ref{fig:applications}b).
For instance, when embodying a dragon, characterized by heavy, powerful movements, a designer can adjust the exoskeleton’s 6 actuated DoFs with the \textit{resist} function to increase resistance, simulating large, forceful wing beats. Conversely, for a sparrow, with its fast and light motions, the designer can reduce resistance or even employ the \textit{amplify} feature to emphasize the sparrow's agility. Through iterative testing and adjustment of the \textit{resist} and \textit{amplify} parameters, we can fine-tune the kinesthetic feedback, creating avatar-specific haptic experiences.

\paragraphB{Collaborative body-actuated play}
Body-actuated play~\cite{patibanda_2022} leverages body-actuating technologies for novel kinds of creative, shared bodily experiences. Using \toolkit, we developed a collaborative game that demonstrates how designers can craft interactions between multiple players each wearing an exoskeleton. In this game, one player ~(P1) controls the body movements of another player~(P2). The  goal is to solve a physical color sorting game by making P2 sort as many rings correctly as time permits. 
P1 is wearing a 3-DoF exoskeleton arm with sensing capabilities that captures her movements. P2 wears a fully actuated 3-DoF exoskeleton that mirrors P1's motion in real time~
(see Figure~\ref{fig:applications}c). This behavior was rapidly implemented with the toolkit’s \textit{mirror} function, which supports customizable body remapping strategies in few lines of code. 
To make the game more demanding for advanced players, we developed a second level that adjusted how the transferred motions are scaled by changing one parameter in the software. Finally, we implemented a third level, that goes beyond a direct 1:1 mapping of body parts and instead maps P1's sideways shoulder motions onto P2's elbow motions, and P1's elbow motions onto P2's shoulder joint to challenge coordination. 

\paragraphB{Enhanced artistic performances}
Exoskeletons also offer interesting applications for artistic performances. Using \toolkit, we developed an arm exoskeleton that artistically modifies a dancer’s movements in real time. The dancer defines and executes the overall motion of the arms, while the exoskeleton enhances the style of the motion to be more jerky and robot-like~(see Figure~\ref{fig:applications}d).
By leveraging the pre-implemented \textit{jerk} function, the designer can introduce variability in the dancer’s movements, making them more abrupt. 
Designers can easily adjust jerk parameters, such as amplitude, frequency, and velocity, to suit the performance, from subtle modifications to exaggerated effects. 
Moreover, the toolkit allows designers to easily implement specific body gestures that serve as triggers for starting and stopping the motion augmentation. We implemented an ``arm stretched out'' gesture~(elbow is at 0 degree angle) for triggering the motion style modulation, ensuring the augmentation aligns with the dancer's intent and preferences.


\begin{figure*}[t]
    \centering
    \includegraphics[width=\linewidth]{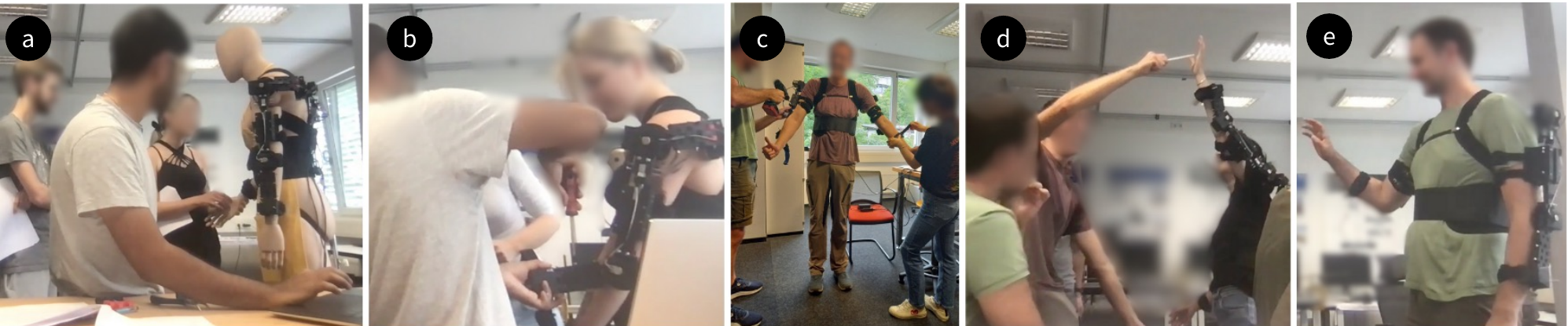}
    \caption{Workshop participants first familiarized themselves with the provided prototype~(a), put it on and collaboratively adjusted the link sizes~(b, c), and experienced the provided functions on their bodies~(d, e).}
    \label{fig:workshop}
    \Description{(a) Participants standing around a mannequin wearing the exoskeleton prototype. (b) One participant wears the exoskeleton while another participant screws some chain segments at the lower arm to adjust the size to the user's dimension. (c) A similar scene with different participants. (d) One participant hands over an object to another participant who wears an exoskeleton. (e) Participant waves.}
\end{figure*}

\subsection{Usage Study 1: Experiencing ExoKit}
The objective of the first study was to gain insights into \toolkit's utility, understand how it aligns with the key design requirements, and identify opportunities for improvement. 

\subsubsection{Method}
To address this objective, we conducted a usage study with six participants in two workshop sessions. The workshop setting has been proven useful to evaluate toolkits in prior work~(e.g.,~\cite{pfeiffer_2016,sabnis_2023,lei_2022}). Each workshop combined walkthrough demonstrations~(cf.,~\cite{ledo_2018}), where participants are presented with ExoKit's functionalities, with phases of free exploration and discussion. 

\paragraphB{Participants} The first workshop was conducted with three participants experienced in interaction design, representing our target group of novice roboticists: a postdoctoral researcher experienced in interaction design for VR/AR~(ID1, 28, male) and two graduate students~(ID2 \& ID3, both 23, female and male) with research experience in interaction design for wearables and first hands-on experience with exoskeletons, respectively. The second session comprised three participants experienced in hardware and fabrication, serving as technical experts to provide feedback on the design and potential improvements: a postdoctoral researcher working on robotics and haptics~(HF1, 32, male), a PhD student primarily experienced in fabrication~(HF2, 24, female), and an undergraduate student with more than ten years of hardware tinkering experience~(HF3, 31, male). 

\paragraphB{Procedure} 
After providing informed consent, we introduced the workshop participants to \toolkit, presenting its conceptual space~(Figure~\ref{fig:conceptual_space}) and physical hardware components~(Section~\ref{sec:hardware}). 
Participants received an exoskeleton consisting of one fully actuated and one sensing arm, as well as a printed manual for all provided interface functions. 
In the first half of each workshop, they familiarized themselves with the functions offered by \toolkit~while the exoskeleton was still attached to a mannequin~(\autoref{fig:workshop}a). 
In the second half, each participant was prompted to wear the prototype and encouraged to adjust its components to fit their body~(\autoref{fig:workshop}b,c). Taking turns, the other group members then used the command-line interface to trigger desired functions~(\autoref{fig:workshop}d,e). 
The workshop was interleaved with rounds of discussion and short brainstorming sessions. 
Sessions were audio- and video-recorded for later analysis. Each session took 3 hours.

\subsubsection{Results}
We collected findings based on our observations of participants' behaviors and statements. We group these results along key design considerations of \toolkit:

\paragraphB{Explorative, creative and iterative design}
Most importantly, in both workshops, the toolkit fostered a playful exploration of human-exoskeleton interactions, especially enabling rapid iterations on design parameters for various functionalities. Notably, the hardware group engaged deeply with extreme parameter settings to test the toolkit's limits, and quickly adopted a creative designer's perspective. For instance, they spontaneously crafted challenges for each other by leveraging motion transfer functions, where unconventional body remappings~(such as mapping a shoulder movement onto the elbow) complicated the fulfillment of the challenge~(\autoref{fig:workshop}d). HF2 later highlighted how this could bear potential to design fun collaborative and creativity-based interactions between users.
Similarly, in the interaction group, the exploration of resistance and amplification, and how these affected the body movements, immediately sparked ideas for possible applications, including a simulated weight-lifting in VR or leveraging them for avatar embodiment. 

In a follow-up brainstorming session, participants proposed more ideas across various domains: Ideas in healthcare included providing adaptive support for weak arms (ID2) and mirroring motions between therapists and patients to aid rehabilitation~(HF2). ID3 envisioned teleoperation scenarios, such as controlling a fatigued arm with the opposite limb during physically demanding tasks like wall painting. ID2 and ID3 also discussed about social applications, like providing hugs over a distance, enhanced with thermal feedback. Ideas for leisure and gaming featured interactive game interfaces and leveraging the exoskeleton to teach dancing or other sports~(HF2). Finally, participants also highlighted \toolkit’s prospective applications for STEM education~(ID2, ID3).

\paragraphB{Programmability and functional abstractions}
Participants found the provided functional abstractions generally easy to understand and quickly grasped the concepts. They further appreciated the offered parameters which allow to rapidly modify the strategies' effects, such as tuning the motion style with different jerk parameters, or experimenting with varying torques applied for the motion effort strategies. This underlines the toolkit's potential to empower its users to leverage complex exoskeleton functionalities encapsulated by the identified functional abstractions. 
Further, participants acknowledged that the provided functional abstractions provide a broad and generic support for various applications. HF3 indicated that for more specialized purposes, it would be advantageous if the designer can define their own augmentation strategies building upon the provided functionalities. This resonates with the different ways of programming offered by \toolkit~(cf. Figure~\ref{fig:programmability}), which facilitate the customization of new interactive behaviors. 
Finally, both groups suggested that for an initial interactive exploration of the provided augmentation strategies, a graphical user interface that visualizes predefined input parameter ranges for different functions would be a promising extension\footnote{Of note, the GUI presented in \autoref{subsec:interfaces} was implemented in response to this feedback.}.

\paragraphB{Wearability}
Both groups successfully adapted the exoskeleton’s size by adjusting the link lengths to their body sizes, demonstrating how the modular components can enhance ergonomics. The interaction group expressed themselves particularly positive about the wearing comfort and lightweight design. Especially ID2, who was familiar with other open-source exoskeleton prototypes, acknowledged that the prototype built with \toolkit~enhanced comfort. Finally, the hardware group offered suggestions to further improve wearability, as they criticized that attaching and detaching the prototype currently requires a second person. For instance, they suggested using more easily attachable backpack straps instead of the postural corrector, and sewing the joint elements directly onto some fabric to further enhance stability.

\paragraphB{Customizability of hardware}
To further customize \toolkit~to the versatile needs of HCI researchers, both groups emphasized the importance of being able to integrate additional sensing and actuation mechanisms, such as thermal feedback or vibration. This requirement is already supported by \toolkit's current infrastructure which is intentionally built on Arduino and thereof naturally compatible with a wide set of commonly used, relevant electronics. Additionally, HF3, who has a strong background in hobby tinkering, requested support for low-cost joint modules as the currently supported Dynamixel motors are comparatively expensive for private use beyond research.


\begin{figure*}[t]
    \centering
    \includegraphics[width=\linewidth]{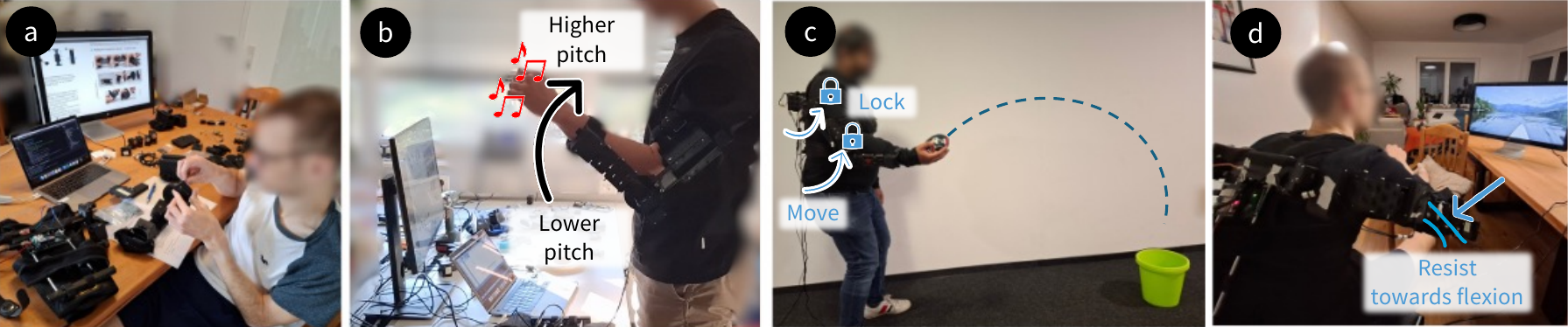}
    \caption{Participants used ExoKit's software and hardware components~(a) to implement an application of their choice: 
    P1 created a sound synthesizer~(b), P2 an exoskeleton that teaches how to do an underhand toss~(c), and P3 a rowing application~(d).}
    \label{fig:solodiy}
    \Description{(a) A participant sits in front of a table. Around him, ExoKit's hardware components are spread out. Two screens in front of him display the user manual and a programming interface. (b)-(d) is a visual representation of the built applications which are described in subsection 6.3.}
\end{figure*}

\subsection{Usage Study 2: Prototyping with ExoKit}

The goal of the second study was to gain rich qualitative insights into how users approach the toolkit to build an application, their workflows, the components they would use and encountered challenges.

\subsubsection{Method}
Inspired by Lei et al.~\cite{lei_2022}, we conducted a usage study in which users engaged with ExoKit over a longer period of time. 

\paragraphB{Participants} We recruited three participants from the target group, all with an interest in learning about human-exoskeleton interactions: an undergraduate student with some theoretical knowledge of human-exoskeleton interaction who self-reported basic electronics and programming skills~(P1, 22, male), a graduate student with some background in design and fabrication of actuated systems, good electronics and basic programming skills ~(P2, 28, male), and a PhD student in HCI interested in sports and accessibility without any prior robotics experience but basic electronics and good programming skills (P3, 26, male).

\paragraphB{Procedure} Participants worked in a quiet space equipped with ExoKit (software, tools, user manual) and pre-fabricated hardware modules~(\autoref{fig:solodiy}a). In response to the feedback from usage study~1, we additionally introduced a GUI to facilitate first-time exploration, replacing the command-line interface~(cf., \autoref{subsec:interfaces}).
Participants were tasked with creating a functional human-exoskeleton interaction for an application of their choice, with no time constraints. The only requirement was to present a live demonstration at the end. 
After a brief introduction, we taught participants how to get the basic setup (comprising an Arduino, a connected motor and a programming interface) running. Next, participants familiarized themselves with the offered functions with the help of the GUI before starting to develop their own applications.
To gather in-depth qualitative insights, we conducted observations, took notes, and engaged in short conversations with the participants from time to time to assess their progress. 
The study concluded with semi-structured interviews, which were audio-recorded and transcribed for analysis.

\subsubsection{Results}
We thematically analyzed our notes and interview transcripts and generated the following themes: 

\paragraphB{Implemented applications}

{\renewcommand{\arraystretch}{1.4}
\begin{table*}[t]
    \centering  
    \begin{tabular}{p{0.4cm}  >{\columncolor[gray]{0.95}}p{2.5cm}  p{6.5cm}  >{\columncolor[gray]{0.95}}p{6.5cm}}
        \textbf{ID} & \textbf{Application} &  \textbf{Programming Interface \& Functions} & \textbf{Hardware Components} \\
        \hline
        \hline
        P1 & Sound synthesizer & Arduino library and Processing application; continuously streaming joint angle data \hspace*{\fill}(2.5 h) & 1 active elbow joint module, links, cuffs \hspace*{\fill}(20 min) \\
        \hline
        P2 & Teacher for learning a toss & Processing library; a sequence of joint and arm-wise \textit{moveTo()} \& \textit{lock()} functions \hspace*{\fill}(3.5 h) & 2 active (elbow \& shoulder side) and 1 passive shoulder back joint module, links, cuffs, back rail \& posture corrector \hspace*{\fill}(1.5 h) \\
        \hline
        P3 & Rowing trainer & Arduino library; a sequence of \textit{resist()} functions with variable torques, executed in parallel on both joints \hspace*{\fill}(40 min) & 2 active elbow \& 4 passive shoulder joint modules, links, cuffs, back rail \& posture corrector \hspace*{\fill}(3 h) \\
    \end{tabular}
    \caption{An overview of the applications built in the second usage study, including the software and hardware components that the participants decided to use. Within the indicated time, participants familiarized themselves with the library functions, implemented the software, and learned to assemble the exoskeleton prototype.}
    \Description{P1 built a sound synthesizer. He leveraged the Arduino library and built a Processing application. He took 2.5 hours for the software implementation. For the hardware assemble, he took 20 minutes and leveraged an active elbow joint module, links and cuffs. P2 created a wearable teaching device for learning a toss. He used the Processing library to implement a sequence of joint- and arm-wise moveTo() and lock() functions. He took 3.5 hours for the software. He assembled the hardware within 1.5 hours, leveraging an active elbow and shoulder side joint module and one passive module attached to the back of the shoulder, links, cuffs, and the back rail attached to the posture corrector. P3 built a rowing trainer. He used the Arduino library to implement a sequence of resist() functions with variable torques that are executed in parallel on both active elbow joints. For the software, he took 40 minutes. Buildung the hardware took 3 hours. He used 2 active elbow joint modules, 4 passive joint modules for the shoulder, links, cuffs, and the back rail attached to the posture corrector.}
    \label{tab:placeholder}
\end{table*}
}
P1 created a \textbf{sound synthesizer} where the position of the elbow joint is transformed into a sound in real-time (\autoref{fig:solodiy}b). 
Similar to a theremin, the farther the lower arm moves towards flexion, the higher the pitch becomes. The idea was inspired by the user manual which provides application ideas to inspire novel users, and resonated with P1's interests as he has \textit{``some colleagues [who] built an application to use novel input devices to create sounds''} .
P1 used the Arduino library to continuously stream motion data to an external application. He implemented a small Processing application to parse and sonify the incoming sensor data using Processing's sound library\footnote{Processing's sound library: \url{https://processing.org/reference/libraries/sound/index.html}, last accessed December 3, 2024}. 

P2 designed a \textbf{device that teaches him to do an underhand toss}, as he was \textit{``not particularly athletic and I thought that an exosuit could teach me to kind of perform this underhand toss''} (\autoref{fig:solodiy}c). Using two motors, P2 actuated the right shoulder and elbow to support flexion-extension.
With the help of the Processing library, P2 implemented the required motion sequence: both actuated joints first move to a neutral position, then the exoskeleton performs a throwing motion by moving the joints towards the targeted angles, followed by a brief lock to keep this position. 

P3 created a \textbf{training device for rowing}, using an exoskeleton with two actuated elbow joints to provide resistance during arm movements. P3 implemented a personalised training plan, in which the provided resistance increased in 5-second intervals to a user-defined maximum before gradually decreasing again. For a more realistic feeling, the resistance was only provided during elbow flexion. P1 synchronized his rowing movements with an online first-person rowing video for demonstration (\autoref{fig:solodiy}d).

\autoref{tab:placeholder} provides an overview of the software and hardware components that the participants used and the time they took to build the application.

Participants reflected on the learning curve, which could be expected given the multitude of modular software and hardware components with which the participants had to become familiar first. For instance, P3 stated: \textit{``I would have a huge advantage if I do it again because I mean at the end it is all very clear and I understand the modularity [\ldots] but it was just a lot to learn''}. 
Of note, in the end, all participants were able to design, build and implement their own exoskeleton interaction.

\paragraphB{Different approaches to developing an application with ExoKit}
Interestingly, the three participants demonstrated three distinct approaches to building applications with ExoKit:  
P3 started by designing the complete hardware, considering \textit{``how large should the exoskeleton be, which joints should it involve, should it actually have all the joints and if yes should they be passive or not be passive''}, before programming the interaction in one go using the Arduino library.  
Contrary, P1 prioritized getting the programming logic right before assembling any hardware. Working only with a minimal hardware setup (a motor attached to the Arduino), he explained \textit{```because while the exo is not fully assembled, I found it easier to actually test the [functionalities] I wanted to use. Especially since I was not that familiar with ExoKit yet, I really enjoyed it was not assembled yet''}. 
P2, being less experienced in interaction design and programming, began by studying tutorials and his own body motions to identify the desired sequence, first drafting it on paper \textit{``to translate this into sort of like a programmatic approach''} and building the corresponding hardware. Afterwards, he deployed an iterative programming approach, gradually increasing the complexity and refining parameters.  
These distinct approaches highlight how ExoKit's modularity provides flexibility to accommodate users with different levels of experience and preferences.  

\paragraphB{The libraries' perceived ease of use depends on user's prior experience}
All participants highlighted that the GUI is \textit{``most ideal for exploration''} (P1) as it requires minimal input: \textit{``you just have to click, you don't have to enter numbers''} (P3). Moreover, the participants emphasized that their first application ideas were inspired during the initial exploration with the GUI, thereby demonstrating its effectiveness in supporting novice users. 
In contrast, we noted that the perceived ease of the Arduino library varied with participants' prior programming experiences, underscoring the importance of providing different programming interfaces. P3, with the most programming experience, found the Arduino library \textit{``easy enough''} for his needs. P1, a less experienced student, mentioned that
it was challenging in the beginning, but he eventually gained \textit{``enough knowledge to work and play with it''}.
For P2, we provided support after observing that he got stuck trying to learn the Arduino API, and pointed out the possibility of starting with a simpler programming interface first. 
P2 concluded that the Processing library aligned better with his skill level: \textit{``If the [GUI] is for people who have no programming experience and if using the Arduino library is for people who have lots of development experience, then I should stick to my strengths which is being in the middle''}.

\paragraphB{Helping users to help themselves}
Our study showed that ExoKit supports users to adopt their own workflows, but also revealed opportunities to improve support when they struggle.  
First, similar to the workshop studies, the participants quickly grasped the provided functional abstractions and explored parameters independently with the GUI, demonstrating the suitability of the abstraction levels and the provided support through the manual.  
Second, to program the interactions, P1 and P3 successfully built on code snippets provided in the manual and tailored them to their needs. P1 and P2 aspired even more examples and detailed code explanations in the manual as this would further ease the learning process. 
Third, P1 and P2, who independently assembled the hardware, emphasized that \textit{``the instruction manual made it pretty straightforward to figure out how it is to be assembled''} (P2), though P1 noted occasional confusion due to similar-looking parts. 
We observed similar moments of confusion for all participants, such as inserting the rigid link into the arm rail first in the wrong orientation. We see potential to reduce such mistakes by integrating more affordances and constraints into the 3D-printed components, mechanically preventing incorrect orientations or connections. Finally, participants recommended to further enhance the assembly instructions through videos (P3), more visuals (P2), or schematic drawings (P1). 
\subsection{Discussion, Limitations \& Future Work}
We now discuss lessons learnt from the implementation of the application cases along with insights from the usage studies. We also discuss limitations of our current implementation and derive implications for future work on exoskeleton toolkits in HCI. 

\paragraphB{Exploring human-exoskeleton interactions}
Our example applications and usage studies demonstrate how \toolkit~facilitates exoskeleton prototyping and design iteration for non-experts, and hence lowers the entry barrier for engaging with exoskeletons. While prototyping the application cases, we intensely leveraged the modular design of the exoskeleton, which allowed us to easily reconfigure prototypes by swapping electronics and functional parts. As demonstrated in the usage studies, the command-line interface and the GUI enabled rapidly exploring functions and iterating on their parameters. It became apparent that the developers needed to fine-tune the parameters of real-time modifications, which modify ongoing user motions, such that the combined motion of both user and exoskeleton felt smooth. Here, too, the provided interfaces proved helpful for rapid iterations. We also noted that the screws required for assembly slowed down prototyping. While other toolkits such as MiuraKit~\cite{cui_2023} 
deploy plug-and-play mechanisms which allow for rapid reconfigurability of the robotic structure, they must be carefully designed in case of \toolkit~to withstand the mechanical stresses on the exoskeleton, a point for future refinement.
Furthermore, \toolkit~opened up new possibilities for exploring human-exoskeleton interactions, beyond the initially anticipated use cases. Both the study participants and the authors came up with novel ideas for interaction designs, such as unconventional body remapping via motion transfer functions. This underlines \toolkit's potential as a platform for creative exploration for the HCI community. 

\paragraphB{Ease of programming interactive behavior}
Our usage studies showed that all users could rapidly grasp and modify the behavior of the augmentation strategies through the command-line interface and GUI.
Our example applications further demonstrate more complex use cases which were implemented with the Arduino firmware.
While the provided functional abstractions present a first essential step to making complex exoskeleton behaviors accessible to novice roboticists, we identified the following functionality to be desirable in future extensions: first, the possibility to control exoskeleton motion in 3D world coordinates, rather than only in joint angle space; second, adding more sophisticated force profiles for motion guidance~\cite{proietti_2016,gasperina_2021}, and third, a record-and-replay mechanism for asynchronous motion transfer that captures motions and replays them at a later point~\cite{proietti_2016,gasperina_2021}. 
We further found that the perceived ease of use of the provided programming interfaces depends on to the user's programming experiences. Drawing on these insights, an additional worthwhile direction to further ease programmability and make the toolkit accessible for even broader audiences is to extend beyond text-based programming to novice-friendly visual programming, such as block-based approaches that could be used instead of the Arduino library.
Finally, a simulation tool could provide further support to the designer, for simulating and testing interactions before deploying them on an exoskeleton. Yet, it is an open question how to most effectively simulate and communicate the effects of body-augmenting systems on the human body.

\paragraphB{Extending support to other body parts}
\toolkit~currently supports up to three active DoFs for each arm, opening up new opportunities for interaction design in the HCI community. However, we are far from supporting the whole body. Extending support to other body parts could unlock new interaction opportunities. For instance, future iterations could add more DoF at the arms, by including internal rotation. Another promising avenues is to extend support to the lower limbs, including the hip and knees. We expect that our fundamental design decisions~(modular components, exchangeable joint types, etc.) will generalize to the lower limbs. However, these pose additional safety challenges, as stronger actuation forces are required and if not done right, these could negatively affect the user's balance. In consequence, future work should also expand the conceptual space to address new augmentation strategies for the lower body, such as gait assistance for leg exoskeletons~\cite{baud_2021}.

\paragraphB{Fabrication \& Cost}
The cost of realizing an exoskeleton with \toolkit~centrally depends on the cost of the motors (between 270 to 430 USD). In comparison, the cost of the fabrication materials is negligible. Relying on 3d printing filament is one factor that makes this technology accessible to a larger audience. While the resulting prototype is sufficient for practical use in early iterations, of note, it also inevitably involves compromises when compared to professional exoskeletons, which are often made from metal structures suited for bearing heavy loads.
Finally, while the overall cost is only a fraction of the price tag of a professional exoskeleton, this may be still too expensive for private users. It is a worthwhile direction to investigate how lower-end, lower-cost motors can be integrated with \toolkit. We invite the community to contribute to the advancement of \toolkit, adapting existing modules and sharing new ones. 

\section{Conclusion}
This paper presented \toolkit, a do-it-yourself toolkit that empowers novice roboticists with basic electronics and programming skills to rapidly prototype interactions for functional lo-fi exoskeletons targeted at the arms. 
\toolkit~features modular hardware components that allow to easily reconfigure its active degrees of freedom, adjust component's dimensions to accommodate various body sizes, and safety mechanisms. We conceptually identified relevant high-level augmentation strategies and provide them as functional abstractions that simplify the programming of interactive behaviors. These functions are readily accessible and customizable through a command-line interface, GUI, Processing library, and Arduino firmware. 
Through application cases and two usage studies, we demonstrated \toolkit's potential to ease the development of human-exoskeleton interactions and support creative exploration and rapid iteration in early-stage interaction design. We hope that this work will inspire HCI researchers to explore the emerging field of human-exoskeleton interaction and unlock its potential for innovative applications.
 
\begin{acks}
    We thank all participants of our usage studies and express our particular gratitude to Ata Otaran for his feedback. We also thank the reviewers for their valuable comments.
\end{acks}

\bibliographystyle{ACM-Reference-Format}
\bibliography{revised_references}

\appendix
\section{Software Implementation} \label{subsec:implementation_sw}
The Arduino firmware was developed with C++ in platformIO. The firmware communicates with the hardware via USB using a serial breakout board and manages the low-level motion control, while the Dynamixel motors internally handle the hardware-level motor control.
In a first step, users configure the exoskeleton in the firmware with the \texttt{ExoskeletonBuilder}, which follows a builder pattern to add configured joints and define related hardware components and joint ranges, followed by a calibration step which determines the absolute zero-position for each actuated or sensing only joint. These configurations can be persisted in the Arduino's EEPROM for retrieval between microcontroller restarts.

In our event-based architecture, interactive behavior is modeled with \texttt{Condition} and \texttt{Action} classes. Conditions serve as runtime triggers that control the flow of actions, evaluating to true or false based on the exoskeleton’s current state. This evaluation is checked in a control loop that runs at a fixed frequency. Actions encapsulate basic functions and augmentation strategies~(cf., Section~\ref{sec:concepts}). They can be combined sequentially or in parallel either for individual joints, a set of joints, or the arm for complex control flows. These are managed using \texttt{ActionBuilder}, which abstracts the sequentializion and parallelization of actions through another builder pattern, and provides an intuitive programming interface to the user.
Additionally, custom commands can be registered to trigger augmentation strategies from the outside through a serial console. The provided command-line interface, GUI, and Processing library build on this capability. 

The algorithms for implementing the functional abstraction are provided in the supplemental material. For the full implementation, please refer to the project's GitHub repository\footnote{\url{https://github.com/HCI-Lab-Saarland/ExoKit}}

\section{Hardware Implementation}\label{subsec:implementation_hw}
We designed the hardware components with Autodesk Fusion 360 and fabricated the 3D-printed parts with an Ultimaker S5 printer, using ABS filament. We recommend 100\% infill with an octet infill pattern for enhanced stability~\cite{pernet_2022} as well as a normal support structure and 2 mm wall thickness. The modular 3D-printed components are assembled with screws, nuts, and bearings, enabling easy adjustment.
The exoskeleton is held in place through a 3D-printed back rod, which is sewn onto a posture corrector. Two additional aluminium rods at the back further stabilize the system. 

\toolkit~contains only commercially available electronics. It supports the four motors of the Dynamixel XM series that we chose because of their balance between power and cost-efficiency. For sensing, we use Adafruit-based sensors, including load cells with HX711 ADC chips and a panel mount 10K potentiometer for angle tracking. The motors interface with the Dynamixel Shield for Arduino, while an Arduino Mega manages the communication with both the motors and sensors. A Sparkfun serial basic breakout~(CH340C) facilitates serial communication with the shield. The motors can be powered with 12 Volt using the Dynamixel power supplies or a portable power bank. 
For the portable power bank, we leveraged a Zeee Lipo Battery (3300mAh, 14.8V 50C) in a 4S1P configuration (348 grams, 131$\times$43.5$\times$29 mm. We connected the power supply to three voltage converters (12V, 5A, 60 grams) to power the Dynamixel shield and two SMPS2Dynamixel modules inserted into the daisy-chained motors to supply all 6 motors with power. Another voltage converter (5V, 3A, 50 grams) is used to power the Arduino. We put the converters into a box attached to the back of the posture corrector. This resulting setup is completely mobile (see minute 1:30 of the supplemental video).

\end{document}